\documentclass[fleqn,12pt,twoside]{article}
\usepackage[headings]{espcrc1}

  \def\be{\begin{equation}}      
  \def\ee{\end{equation}}        \def\beq{\begin{eqnarray}}
          \def\eeq{\end{eqnarray}}
       \def\m{\multicolumn}
             \def\ra{\rightarrow}
\readRCS
$Id: espcrc1.tex,v 1.2 2004/02/24 11:22:11 spepping Exp $
\usepackage{graphicx}
\usepackage[figuresright]{rotating}

\newcommand{\AmS}{{\protect\the\textfont2
  A\kern-.1667em\lower.5ex\hbox{M}\kern-.125emS}}

     \def\beq{\begin{eqnarray}}
  \def\eeq{\end{eqnarray}}    
  \def\be{\begin{equation}}  \def\ee{\end{equation}}

\title{Decay rates of Quarkonia in the NRQCD formalism}

\author{J N Pandya,\address{Applied Physics Department, Faculty of Technology and Engineering,\\
 M S University of Baroda, Vadodara, Gujarat-390 001, INDIA.}
 Ajay Kumar Rai$^{a,}$\address[MCSD]{Department of Physics, Sardar Patel University,
Vallabh Vidyanagar, \\ Gujarat-388 120, INDIA.} and
           P C Vinodkummar  \addressmark[MCSD]}

\runtitle{Decay rates of Quarkonia in the NRQCD formalism}
\runauthor{J N Pandya et al.}

\begin{document}
\maketitle
\begin{abstract}
Decay rates of $c\bar c$ and $b\bar b$ mesons have been studied
within the NRQCD formalism. The basic parameters of the formalism
have been obtained from different potential schemes studied for the
quarkonia spectra. The present results are compared with other
potential model results with and without correction terms proposed
through hard gluon processes involved in the decay.
\end{abstract}
\section{Introduction}
The mesonic states are not only identified with their masses but
also with their leptonic and other decay rates. So, one of the
tests for the success of any theoretical model for mesons is the
correct prediction of their decay rates. Many theoretical models
predict the masses correctly but overestimate the decay rates
\cite{BuchmullerTye1981,JNPandya2001,Martin1980,Quiggrosner1977,Eichten1978,AKRai2002}.
For better estimations with reference to the experimental values,
various corrections due to radiative processes, higher order QCD
contributions etc were suggested \cite{Hafsakhan1996}. In this
context, the NRQCD formalism is found to provide systematic
treatment of the perturbative and non-perturbative components of
QCD at the hadronic scale \cite{Bodwin}. For the present study, we
employ phenomenological potential scheme for the bound states of
heavy quarkonia and the resulting parameters and wave functions to
study the decay properties.

\section{Nonrelativistic Treatment for Heavy Quarks}
For the heavy-heavy bound state systems such $c \bar c$, $b \bar
b$ as, we consider a nonrelativistic Hamiltonian given by
\cite{AKRai2002}
\begin{equation}
\label{eq:nlham}H=M+\frac{p^2}{2M_1}+V(r), \ \ where \ \ \ M = m_Q
+ m_{\bar Q} , \ \ \  \ {\rm and} \   \  \ \ M_1=\frac{m_Q \
m_{\bar Q}}{m_Q + m_{\bar Q}}\end{equation} $m_Q$ and $m_{\bar Q}$
are the mass parameters, $p$ is the relative momentum of each
quark and $V(r)$ is the quark antiquark potential given by
\cite{AKRai2002} \be V(r)=\frac{-\alpha_c}{r} + A r^\nu; \ \
\alpha_c=\frac{4}{3} \  \alpha_s \label{eq:405}\ee
 We employ hydrogenic trial wave function within the variational
scheme such that the expectation value of the Hamiltonian for the
ground state given by
 \begin{equation} \label{eq:toteg}E(\mu,\nu )=M+\frac{1}{8} \
 \frac{\mu^2}{M_1}+\frac{1}{2}\left(- \mu \
\alpha_c + A \ \frac{\Gamma(\nu+3)} {\mu^{\nu}}\right)
\end{equation}
is minimized to determine the wave function parameter $\mu$. The
spin average mass $(M_{SA})$ of the system is obtained using
Eqn(\ref{eq:toteg}). The results for $c \bar c$ and $b \bar b$
systems are tabulated in Table-1 along with the square of the
radial wave functions at the origin for different choices of
$\nu$. The predictions of the other contemporary potential models
are also listed. The mass parameters $ m_b = 4.66 \ GeV$ $,
m_c=1.31 \ GeV$ and the mass difference between the pseudoscalar
(P) and vector meson (V) due to the chromomagnetic hyperfine
interaction is obtained as described in  Ref \cite{AKRai2002}. The
Potential parameter $A$ of $CPP_{\nu A}$ is fixed to get the
ground state masses of the quarkonia.

\section{Decay rates of $c \bar c$ and $b \bar b$ mesons
in NRQCD formalism}

The decay rates of the heavy quarkonium states into light hadrons
and into photons and pairs of leptons are among the earliest
applications of perturbative quantum chromodynamics (QCD)
\cite{app,de,Barbieri1979}. The decay rates of the meson are
factorized into a short-distance part that is related to the
annihilation rate of the heavy quark and antiquark and a
long-distance part containing all nonperturbative effects of the
QCD.
\begin{table}
\caption{Theoretical predictions of the masses (in GeV) and
$|R(0)|^2$ of $c\bar c$ and $b\bar b$ systems.}\label{tab:32}
\begin{tabular}{lllccccll}
\hline & Models &  $\alpha_s $& \m{2}{c}{$M_P$}&
\m{2}{c}{$M_V$} &$M_{SA}$&$| R_{cw}(0)|^2$ \\
&&&Theory&Expt\cite{particle2002}&Theory&Expt\cite{particle2002}&GeV&$GeV^3$\\
\hline
&ERHM\cite{JNPandya2001}&0.356&2.985&2.9804&3.096&3.097&3.068&0.556\\
&BT\cite{BuchmullerTye1981}$$&0.360&2.980&$\pm$&3.097&&3.067&0.810\\
$c \bar c$&PL(Martin)\cite{Martin1980}&0.430&2.980&0.0012&3.097&&3.067&0.999\\
&Log\cite{Quiggrosner1977}&0.370&2.980&&3.097&&3.067&0.815\\
&Cornell\cite{Eichten1978}&0.310&2.980&&3.097&&3.067&1.454\\
&$CPP_{\nu A}$\cite{AKRai2002}$\nu = $ 0.5&0.300&2.997&&3.092&&3.068&0.609\\
&\hspace{1.6cm} $\nu = $ 0.7&0.300&2.973&&3.099&&3.068&0.814\\
&\hspace{1.6cm} $\nu = $ 1.0&0.300&2.940&&3.111&&3.068&1.099\\
&\hspace{1.6cm} $\nu = $ 1.5&0.300&2.893&&3.127&&3.068&1.509\\
\hline
&ERHM\cite{JNPandya2001}&0.241&9.452&&9.464&&9.461&4.990\\
&BT\cite{BuchmullerTye1981}&0.241&9.377&9.300&9.464&9.4603&9.440&6.477\\
$b \bar b$&PL(Martin)\cite{Martin1980}&0.270&9.398&$\pm$&9.462&$\pm$&9.446&4.591\\
&LOG\cite{Quiggrosner1977}&0.245&9.395&0.002&9.460&0.00026&9.444&4.916\\
&Cornell\cite{Eichten1978}&0.217&9.335&$\pm$&9.476&&9.441&14.05\\
&$CPP_{\nu A}$\cite{AKRai2002}$\nu = $ 0.5&0.233&9.445&0.002&9.457&&9.453&3.909\\
&\hspace{1.6cm} $\nu = $ 0.7&0.233&9.442&&9.457&&9.453&4.779\\
&\hspace{1.6cm} $\nu = $ 1.0&0.233&9.440&&9.458&&9.453&5.988\\
&\hspace{1.6cm} $\nu = $ 1.5&0.233&9.436&&9.459&&9.453&7.728\\
\hline
\end{tabular}
\end{table}

\begin{table}
\caption{$0^{-+} \rightarrow \gamma \ \gamma$ and $1^{-\ -}
\rightarrow l^+ \ l^-$ decay rates (in keV) of $c \bar c$ and $b\bar
b$ mesons.} \label{tab:34}
\begin{tabular}{clcccccc}
\hline
&Models&\m{3}{c}{$0^{-+} \rightarrow \gamma \ \gamma $}
&\m{3}{c}{$1^{-\ -} \rightarrow l^+ \ l^-$}\\
&&$\Gamma_o$&$\Gamma_{NRQCD}$&$\Gamma_{Expt.}$\cite{particle2002} &
$\Gamma_{VW}$&$\Gamma_{NRQCD}$&$\Gamma_{Expt.}$\cite{particle2002}\\
\hline &ERHM\cite{JNPandya2001}&7.67&4.074&&5.55&3.655&\\
&BT\cite{BuchmullerTye1981}&11.19&7.143&7.0$^{+ \ 1.0}_{- \ 0.9}$&8.31&5.357&5.40 $\pm$ 0.15\\
&PL(Martin)\cite{Martin1980}&13.81&9.784&&9.96&7.338&$\pm $0.07\\
&Log\cite{Quiggrosner1977}&11.26&7.209&&8.13 &5.407&\\
$c\bar c$&Cornell\cite{Eichten1978}&20.09&17.18&&14.50&12.885&\\
&$CPP_{\nu A}$ \cite{AKRai2002}$\nu = $ 0.5&8.356&4.674&&6.087&3.506&\\
&\hspace{1.6cm} $\nu = $ 0.7&11.280&7.221&&8.115&5.416&\\
&\hspace{1.6cm} $\nu = $ 1.0&15.399&11.323&&10.911&8.493&\\
&\hspace{1.6cm} $\nu = $ 1.5&21.491&18.209&&14.899&13.657&\\
\hline
&ERHM\cite{JNPandya2001}&0.440&0.285&&1.330&1.034&\\
&BT\cite{BuchmullerTye1981}&0.569&0.513&&1.717&1.775&\\
&PL(Martin)\cite{Martin1980} &0.408&0.394&0.364\cite{Bodwin}&1.216&1.412&1.314 $\pm$ 0.029\\
&Log\cite{Quiggrosner1977}&0.437&0.387&&1.303&1.350&\\
$b\bar b$&Cornell\cite{Eichten1978}&1.258&1.627&&3.719&5.253&\\
&$CPP_{\nu A}$\cite{AKRai2002}$\nu = $ 0.5&0.346&0.239&&1.056&0.844&\\
&\hspace{1.6cm} $\nu = $ 0.7&0.423&0.289&&1.266&1.022&\\
&\hspace{1.6cm} $\nu = $ 1.0&0.530&0.356&&1.586&1.264&\\
&\hspace{1.6cm} $\nu = $ 1.5&0.684&0.451&&2.047&1.604&\\
\hline
\end{tabular}
\end{table}
The short-distance factor calculated in terms of the running
coupling constant $\alpha_s(M)$ of QCD was evaluated at the scale
of the heavy-quark mass $M$, while the long-distance factor was
expressed in terms of the meson's nonrelativistic wave function,
or its derivatives, evaluated at origin. We study the di-gamma
decay of $^1 S_0$ state and the leptonic decay of $1^{- -}$ state
using the conventional Van-Royen Weisskopf formula
\cite{Vanroyenaweissskopf} as well as using the NRQCD formulism
\cite{Bodwin2002}. The NRQCD factorization expressions for the
decay rates are given by \cite{Bodwin2002}
 \be \label{eq:sogma}
\Gamma(^1 S_0 \rightarrow \gamma \gamma)=\frac{F_{\gamma
\gamma}(^1 S_0)}{m^2_q} X+\frac{G_{\gamma \gamma}(^1 S_0)}{m^4_q}
Y; \ \Gamma(^3 S_1 \rightarrow e^+e^-)=\frac{F_{ee}(^3
S_1)}{m^2_q} X+\frac{G_{ee}(^3 S_1)}{m^4_q} Y \ee
 The long distance coefficients $X$ and $Y$ are the NRQCD matrix elements
for the decay. The vacuum saturation approximations allowed the
matrix elements of the four fermion operators to be expressed in
terms of renormalized  wave function parameters \cite{Bodwin}
 \beq X &=& \langle^1 S_0|0_1(^1 S_0)|^1 S_0 \rangle = \frac{2}{3
\pi}|R_{cw}(0)|^2[1+O(v^4)],\cr && \ \ Y=\langle ^1 S_0|P_1(^1
S_0)|^1 S_0 \rangle
 = -\frac{2}{3 \pi}|R_{cw} \bigtriangledown^2 R_{cw}|[1+O(v^4)] \eeq

The short distance coefficients F's and G's computed of the order
of $\alpha_s^2$ and $\alpha_s^3$ as \cite{Bodwin2002}

\be F_{\gamma \gamma}(^1 S_0)=2 \pi Q^4 \alpha^2
\left[1+\left(\frac{\pi^2}{4}-5 \right) C_F \frac{\alpha_s}{\pi}
\right],\ \ \ G_{\gamma \gamma}(^1S_0)=- \frac{8 \pi Q^4}{3}
\alpha^2 \ee

\beq F_{ee}(^3 S_1) &=& \frac{2 \pi Q^2 \alpha^2}{3} \left\{ 1- 4
C_F \frac{\alpha_s(m)}{\pi} \right. 
\left. +\left[-117.46+0.82n_f+\frac{140 \pi^2}{27}
ln(\frac{2m}{\mu_A})\right](\frac{\alpha_s}{\pi})^2 \right\}, \ \cr
&& G_{ee}(^3 S_1)=- \frac{8 \pi Q^2}{9} \alpha^2, \ \ {\sf Where} \
\ C_F=(N_c^2-1)/(2N_c)=4/3. \eeq

Using the relevant parameters from Table-1 . We compute the decay
rates using conventional Van-Royen Weisskopf formula for $\Gamma_0$
$\&$ $\Gamma_{VW}$ as well as using NRQCD expressions. Here $ N_c=3$
is the numbers of colour and $\alpha$ is the electromagnetic
coupling constant. The results are listed in Table-2.
\section{Summary and Conclusion}
The radial wave functions at zero separation and of $c \bar c$ and
$b \bar b$ systems are obtained in different potential models, and
the decay rates of $0^{-+} \ra \gamma \gamma$ and $1^{--} \ra l^+
l^- $ are computed using the formula of NRQCD formalism. The
results are compared with the value obtained using the
conventional formula ($\Gamma_0$, $\Gamma_{VW}$) as well as with
the respective experimental results. Though the predictions using
conventional formula are far from the experimental results, the
prediction based on NRQCD are found to be in accordance with the
experimental values for most cases. The present study in the
determination of the $S$ wave masses and decay rates of $c\bar c$
and $b\bar b$ systems provide future scopes to study leptonic
decay, light hadron decay, various transition rate and excited
states of these mesonic systems \cite{Brambilla2005}. It can be
concluded that the NRQCD formalism has all the corrective
contributions for the right predictions of the decay rates.

\end{document}